\documentclass[final]{svjour3}
\usepackage[dvipdfmx]{graphicx}
\usepackage{rotating}
\usepackage{amssymb}
\usepackage{mathptmx}
\usepackage[numbers]{natbib}
\usepackage{comment}
\usepackage{color}
\usepackage{wrapfig} 
\usepackage{siunitx} 

\usepackage{lineno}

\makeatletter
\journalname{Journal of Low Temperature Physics}

\bibpunct{}{}{,}{s}{}{,}

\begin{document}

\newcommand{\hdblarrow}{H\makebox[0.9ex][l]{$\downdownarrows$}-}

\title{Performance of TES X-Ray Microcalorimeters Designed for 14.4-keV Solar Axion Search}

\author{Y. Yagi$^{*1,3}$ \and R. Konno$^{1,4}$ \and T. Hayashi$^{3}$ \and \\
K. Tanaka$^{1,3}$ \and N. Y. Yamasaki$^{1,3,4}$ \and K. Mitsuda$^{5}$ \and \\
R. Sato$^{6}$ \and M. Saito$^{7}$ \and  T. Homma $^{6,7}$ \and \\
Y. Nishida$^{8}$ \and S. Mori$^{8}$ \and N. Iyomoto$^{8}$
 \and T. Hara$^{9}$}

\institute{
* \email{
yagi@ac.jaxa.jp
}
\begin{enumerate}
\it
\item{}	Department of Space Astronomy and Astrophysics, Institute of Space and Astronautical Science (ISAS), Japan Aerospace Exploration Agency (JAXA), 3-1-1 Yoshinodai, Chuo-ku, Sagamihara-shi, Kanagawa 252-5210, Japan
\item{}	Astromaterials Science Research Group (ASRG), Institute of Space and Astronautical Science (ISAS), Japan Aerospace Exploration Agency (JAXA), 3-1-1 Yoshinodai, Chuo-ku, Sagamihara-shi, Kanagawa 252-5210, Japan
\item{}	Department of Physics, School of Science, The University of Tokyo, 7-3-1 Hongo, Bunkyo-ku, Tokyo 113-0033, Japan
\item{}	Department of Physics, School of Science, Kitasato University, 1-15-1 Kitasato, Minami-ku, Sagamihara-shi, Kanagawa 252-0373, Japan
\item{}	National Astronomical Observatory of Japan (NAOJ), 2-21-1 Osawa, Mitaka-shi, Tokyo 181-8588, Japan
\item{}	Department of Applied Chemistry, Waseda University, Okubo 3-4-1, Shinjuku-ku, Tokyo 169-8555, Japan
\item{}	Research Organization for Nano \& Life Innovation, Waseda University, 513 Waseda-tsurumaki-cho, Shinjuku-ku, Tokyo 162-0041, Japan
\item{}	Department of Applied Quantum Physics and Nuclear Engineering, Kyushu University, Motooka 744, Nishi-ku, Fukuoka 819-0395, Japan
\item{}	Research Network and Facility Services Division, National Institute for Materials Science (NIMS), 1-2-1 Sengen, Tsukuba, Ibaraki 305-0047, Japan
\end{enumerate}
}

\maketitle

\begin{abstract}
A
$^{57}$Fe
 nucleus
in the solar core
could emit
a 14.4-keV
 monochromatic axion through 
 the
 M1 transition
if a hypothetical elementary particle, axion,  exists to solve the strong CP problem.
Transition edge sensor (TES)
X-ray microcalorimeters can detect 
 such axions very efficiently if they are again converted into photons by a $^{57}$Fe absorber.
 We have
  designed and
 produced 
 a
 dedicated TES array with $^{57}$Fe absorbers for the solar axion search. The iron absorber is set next to the TES, keeping a certain distance to reduce 
 the
 iron-magnetization effect on the spectroscopic performance.
 A gold thermal transfer strap connects them.
A sample pixel irradiated from 
 a
 $^{55}$Fe source detected 698 pulses.
 In contrast to
thermal simulations,
we consider that
 the
 pulses include either
 events produced in an
 iron absorber or gold strap at a fraction dependent on the absorption rate of each material.
Furthermore,
photons deposited on the iron absorber are detected through the strap as intended.
 The identification of all events 
 still needs to be completed.
 However, we successfully operated the TES with 
 the unique design
under iron magnetization for the first time.
\end{abstract}


\section{Introduction of Solar Axion Search and TES Detection Sensitivity}
An axion is a hypothetical elementary particle proposed to solve the strong CP problem in quantum chromodynamics [\cite{PecceiandQuinn77, WeinbergandWilczek78}]. The axion is one of the dark matter candidates required in the current cosmology. 
The
 axion can convert to a photon or vice versa in a certain probability via the axion-photon coupling in an external electric or magnetic field. At the center of the Sun, 
photons of $kT\sim \SI{1.3}{keV}$ [\cite{Turck+93}] could produce axions by the interaction with the magnetic field, and their energy spectrum traces the blackbody radiation [\cite{Bibber+89andHoogenveen+92}].
In addition to this process, magnetic dipole (M1) transitions of some nuclei, which 
can
 be excited thermally, could also produce axions 
via the axion-nucleon coupling.
 $^{57}$Fe nuclei have an M1 transition level of $\SI{14.4}{keV}$, and Moriyama 1995 [\cite{Moriyama95}] calculated the  monochromatic axion flux from the Sun at $\SI{14.4}{keV}$. These monochromatic axions would excite the same 
 $^{57}$Fe
 nuclei in a laboratory on the Earth and would be reconverted to 
14.4-keV
 photons. Moriyama (1995) [\cite{Moriyama95}] also proposed to search for
solar
axions by detecting 
these
photons. 
Moreover,
this method has the advantage that
the detector does not need to be tuned
 to the axion mass, unlike dark-matter axion searches.

 A
silicon-based semiconductor
detector searched the 
14.4-keV
photons from $^{57}$Fe film [\cite{Krcmar+98, Derbin+07, Namba07, Derbin+09, Derbin+11}], and the 
best
upper limit on axion mass of $m_{\rm a} < \SI{145}{eV}$ [\cite{Derbin+11}] is obtained at 95\%
confidence limit
in the KSVZ hadronic axion model [\cite{K79SVZ80}]. Other experiments using 
a
proportional gas chamber to detect 
9.4-keV
photons from $^{83}$Kr gas bounded on the hadronic axion mass of $m_{\rm a} < \SI{12.7}{eV}$ (95\%  C.L.) [\cite{Gavrilyuk15, Gavrilyuk18}]. 
In the $^{57}$Fe film experiment, the branching ratio to emit a 
14.4-keV
 photon is only 10.5\%, and the rest 
are converted to electrons or low-energy X-rays. Moreover, the iron film itself absorbs 
about
80\% of the photons. The combination of 
an
 iron film and 
any silicon-based detectors was not able to
detect the self-absorbed thermal energy from axions in the 
film.
Therefore, the detection efficiency was 
 not more than 10\%.
We came up with 
the
 idea that  
 a
transition edge sensor (TES) X-ray
microcalorimeter 
array
 with 
$^{57}$Fe absorbers
could be a solar axion detector. 
This is because
the 
microcalorimeters
 can detect not only 
14.4-keV
 photons but also all dissipated  energy in the 
 absorbers.
Also, the
excellent
energy resolution of TES microcalorimeters would improve the sensitivity 
for the monochromatic axions, with enough converter mass and exposure time.

\vspace{-4mm}
\begin{figure}[htbp]
\begin{minipage}{0.6\hsize} 
\begin{center}
\includegraphics[width=0.92\linewidth, keepaspectratio]{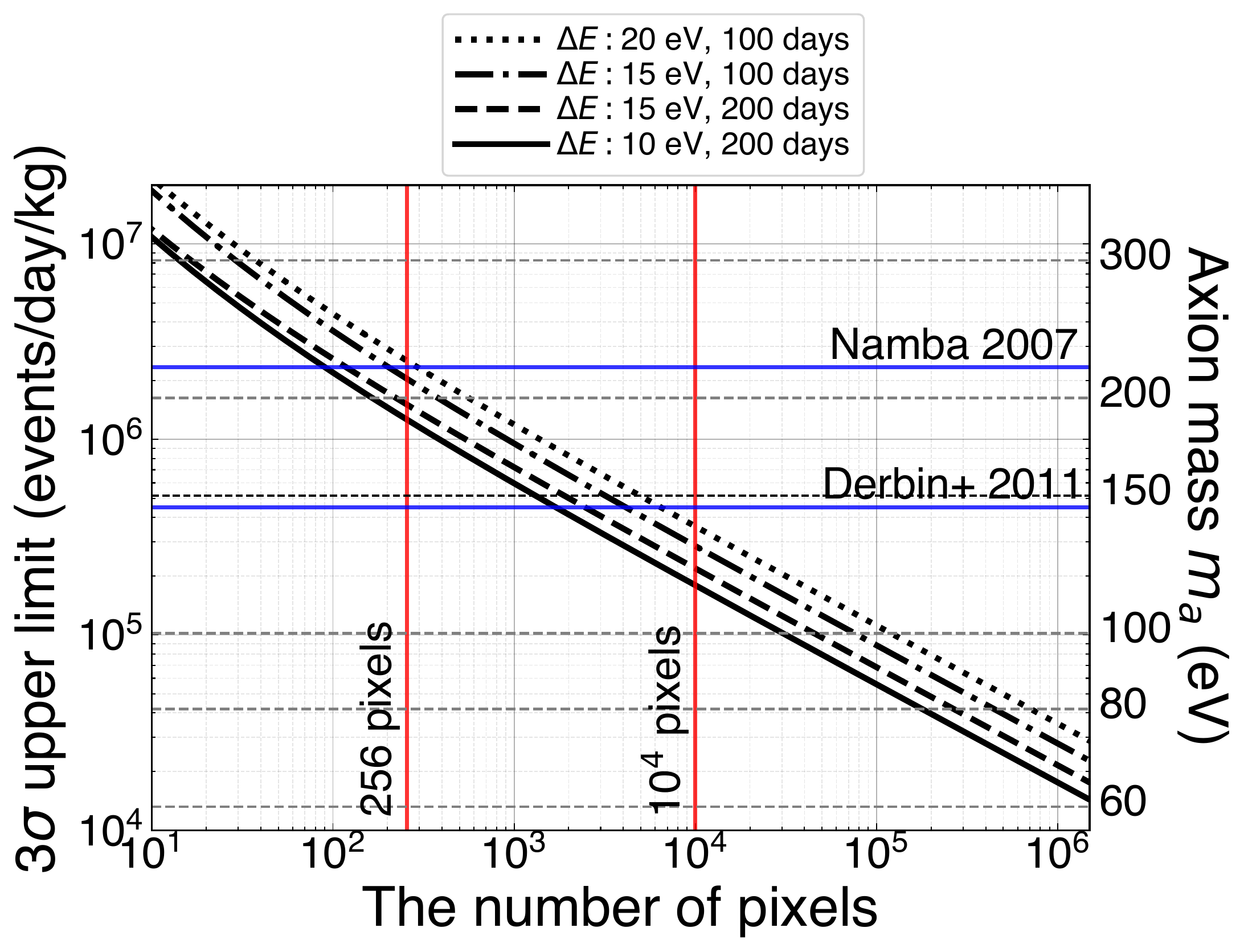}
\end{center}
\label{fig:1}
\end{minipage}
\begin{minipage}{0.37\hsize} 
\begin{center}
\caption{The $3\sigma$ detection limits as a function of the number of pixels whose $^{57}$Fe absorber size is $100$$\times$$100$$\times$$\SI{10}{\micro m}^3$. The exposure time for more than 100 days using a 256-pixel TES array with less than $\SI{15}{eV}$ energy resolution exceeds the sensitivity of Namba 2007 [\cite{Namba07}]. More than a ten thousand-pixel array can search for the axion mass range much less than $\SI{145}{eV}$ in Derbin et al. 2011 [\cite{Derbin+11}]. (Color figure online.)}
\end{center}
\end{minipage}
\end{figure}

\vspace{-6mm}
We 
 assume
a $^{57}$Fe axion converter size of $100\times100\times\,$$10 \, \si{\micro m}^3$ and 
an
iron
density of $\SI{7.874}{g/cm^3}$. The on-ground background rate of the X-ray spectrometer (XRS) for \textit{Suzaku} observatory using the anti-coincidence detector was $\SI{2.0e-3}{counts/s/cm^2/keV}$  [\cite{Kelley+07}]. We conservatively
 expect
the
background rate of $\SI{1.0e-2}{counts/s/cm^2/keV}$. Our group achieved the full width at half maximum (FWHM) energy resolution of $14.41^{+5.05}_{-4.03} \, \si{eV}$ at $\SI{13.94}{keV}$ [\cite{Muramatsu+17}] and successfully fabricated 
a
64-pixel TES array [\cite{Muramatsu+16}]. 
Figure 1 shows
the estimated  $3\sigma$ detection limits against 
the
number of pixels (i.e., axion converter mass),  with various energy resolutions and exposure times.
A detection limit can be converted into an axion mass limit with Eq. 3 and 7 in [\cite{Moriyama95}]. 
Here, parameter values of $D = 0.808$, $F = 0.462$, $S = 0.5$, and $z = 0.56$ in the hadronic axion [\cite{MateuandPich05}] are used. 
As
a
simultaneous readout of 38 TES pixels via microwave multiplex (MWMUX)  readout system [\cite{Mates+08}] has been demonstrated [\cite{Nakashima+19, Nakashima+20}], 
the 
production of 
a extensive
format array system 
is
promising 
with
our technologies.

\vspace{-2mm}
\section{Microcalorimeter Design and Fabrication}
The energy resolution 
of TES
could degrade if it is operated under a magnetic field made by a ferromagnetic iron absorber. The effect depends on the direction of the field through 
a
TES [\cite{Ishisaki+08}]. The perpendicular magnetic field affects badly, but the horizontal field does
not.
In order to reduce the magnetic field on the TES, we 
located the iron absorber next to the
TES and connected 
them with
a gold thermal 
transfer
 strap, as shown in Fig. 2. We have estimated that the magnetic field becomes less than $\SI{1}{\micro T}$, where 
 we 
were able to
 operate TES microcalorimeters without trouble
 if the distance from 
 the edge of
 a 
 5-$\si{\micro m}$-thickness
 iron is more than $\SI{30}{\micro m}$ [\cite{Konno+20}]. 
 Konno et al. 2020 [\cite{Konno+20}]
showed the possible degradation of energy resolution, from $\SI{10}{eV}$ to $\SI{32}{eV}$, by the position dependence in the iron absorber under the condition
  where
 the conductivity of iron and that of gold
 are $\num{6.3e-2}$ and $\SI{3.7e-2}{W/K/m}$ at $\SI{4}{K}$,
 respectively.
Large enough thermal conductivities of both iron and gold are required.

To form thick iron absorbers, we 
 have been
 developing a recipe
 for
  electroplating
pure iron in 
the
facilities of the Nanotechnology Research Center (NTRC)
at
Waseda University.
 Electroplating is very effective method to produce a small structure 
with
expensive materials like  
a
$^{57}$Fe isotope. Although we need a high density and good thermal conductivity in the absorber, pure iron is rarely used commercially, 
 and
thus standard process has not 
 yet
been established. We made several test 
samples
 by reacting in 
 an
 iron-liquid solution of 
 not expensive
 normal $^{56}$Fe.
 We
 measured 
electric resistances
at room temperature and $\SI{4}{K}$
using the LR-700 AC resistance bridge of Linear Research Inc.
We achieved effective deposition even in low solution concentration, 0.05 mol/L, after adjusting the electric current, the time, and additions in the solution. The thermal conductivities, 
$\num{2.0}$--$\SI{5.4}{W/K/m}$ at $\SI{4}{K}$, calculated from the Wiedemann-Franz law, 
 were
  more than 30 times higher  and more efficient heat transmission than 
the former worst case value, $\SI{6.3e-2}{W/K/m}$, used
in 
the
estimation of the degradation by position dependence [\cite{Konno+20}].

We fabricated
 a test 
device
 of this newly designed TES microcalorimeter in the nanoelectronics clean room at ISAS/JAXA 
in
 a process flow shown in Fig. 3. Firstly, 
 proximity-coupled titanium/gold TES bilayer
  was 
  deposited
   by sputtering, and aluminum 
 leads
 were attached. Secondly, 
 gold
 thermal transfer straps of 2-$\si{\micro m}$ thickness were 
evaporated
by the electron beam physical vapor deposition. Thirdly, iron absorbers of 3-$\si{\micro m}$ thickness were 
 electroplated
  on the electrode seed layer. 
Then, the film was electroplated twice because the target thickness was not reached
the first time.
 The device
 has various distances, 0, 20, 60, 100, 200, 300, and 500 $\si{\micro m}$, 
 between
  an iron-absorber edge
  and
a TES edge to avoid the magnetic field from the iron absorber. Finally, membrane structures of SiN$_{\rm x}$/SiO$_2$ under TES were formed by a deep reactive ion etching. Sixteen pixels out of twenty-four were successfully produced, as shown in Fig. 2. 
We have not yet established the fabrication process completely. 
Some photoresist compositions remained and made 
wrinkles after iron electrodeposition in Fig. 3.
The
lift-off of the seed layers failed in some pixels 
in Fig. 2, {\it left}.
We suspect that the rough gold surface, due to 
the
high deposition rate, caused an adhesion reduction with the seed layer on the straps.

\begin{figure}[htbp]
\begin{minipage}{0.48\hsize} 
\begin{center}
\includegraphics[trim=0mm 0mm 0mm 0mm, width=0.97\linewidth, keepaspectratio]{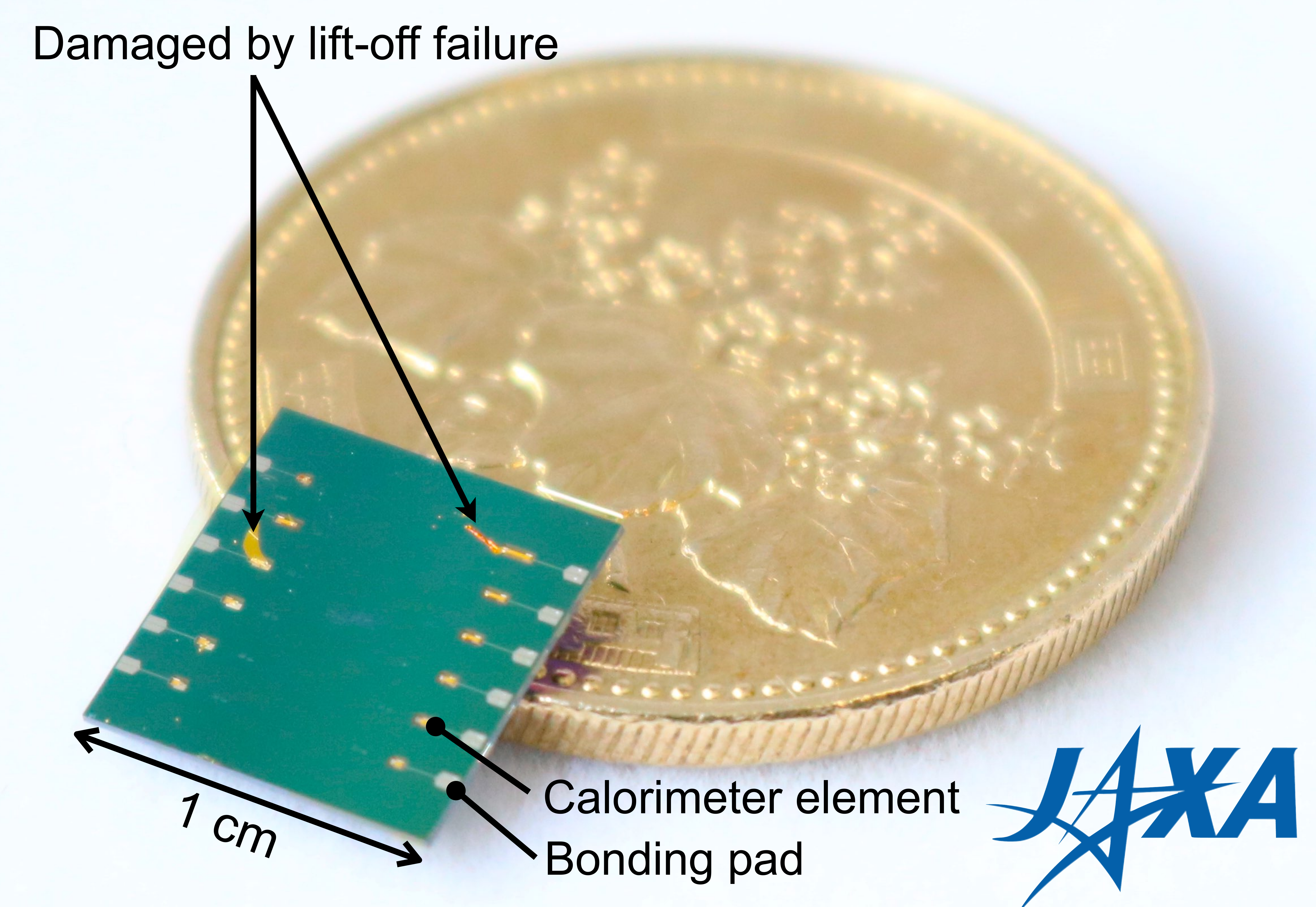}
\end{center}
\end{minipage}
\hspace{1mm}
\begin{minipage}{0.52\hsize} 
\begin{center}
\includegraphics[trim=0mm 0mm 0mm 0mm, width=0.72\linewidth, keepaspectratio]{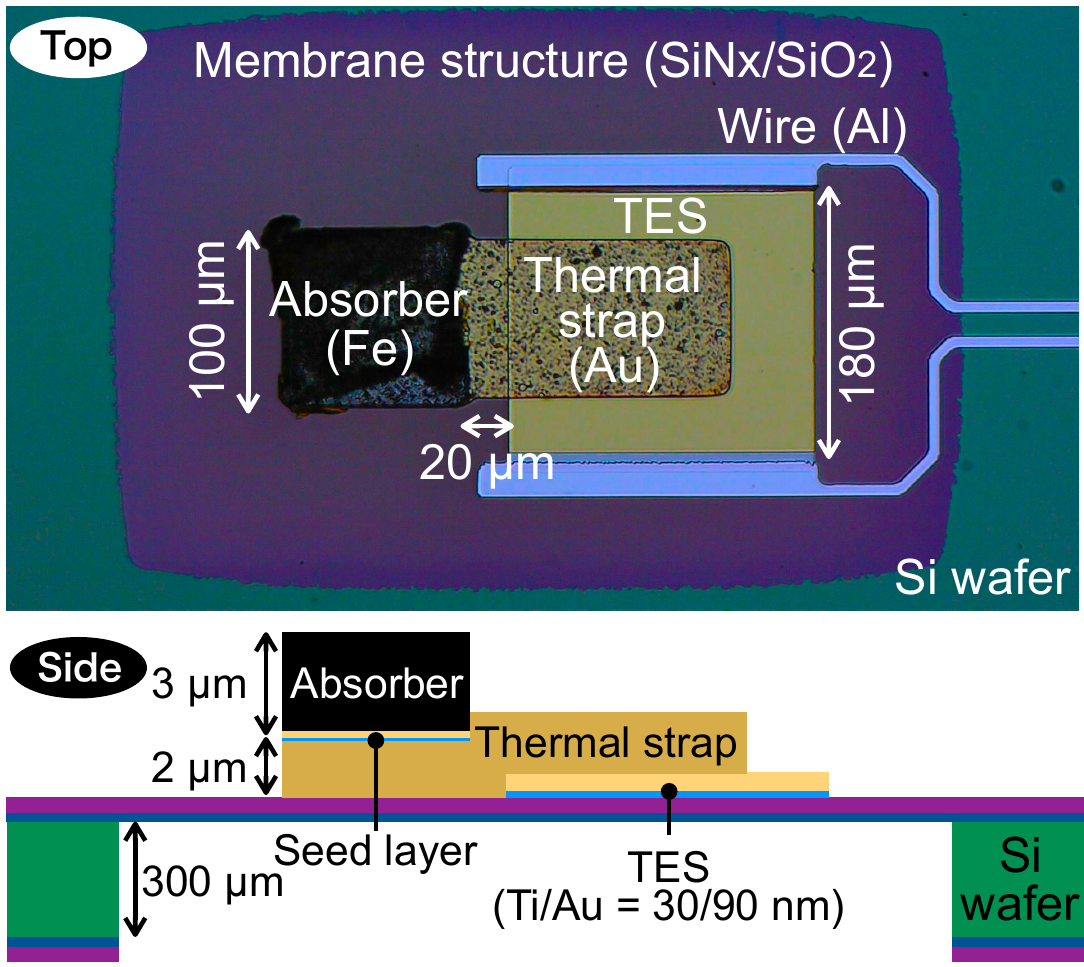}
\end{center}
\end{minipage}
\label{fig:2}
\caption{({\it Left}) Our fabricated chip with 12 elements 
with a Japanese 500-yen coin.
It has various distances, 0, 20, 60, 100, 200, 300, and 500 $\si{\micro m}$, 
between
the iron absorber edge
and
the TES edge
to avoid the magnetic field from the iron
absorber.
Some elements are damaged due to a failure of 
the
lift-off process of seed layers. ({\it Right}) A  top view of a single-pixel TES microcalorimeter, a sample 
element
used in the irradiation test in Sec. 4, and
a side view of its schematic structure.
The size of the TES, iron absorber, and membrane structure are $180$$\times$$\SI{180}{\micro m}^2$, $100$$\times$$\SI{100}{\micro m}^2$, and $300$$\times$$ \SI{460}{\micro m}^2$, 
respectively,
and the thickness of the absorber, thermal strap, and TES (Ti/Au) are $\SI{3}{\micro m}$, $\SI{2}{\micro m}$, and $\SI{30/90}{nm}$, 
respectively.
The
distance between the TES and the iron absorber is $\SI{20}{\micro m}$. (Color figure online.)}
\end{figure}

\vspace{-8mm}
\begin{figure}[htbp]
\begin{center}
\includegraphics[width=0.98\linewidth, keepaspectratio]{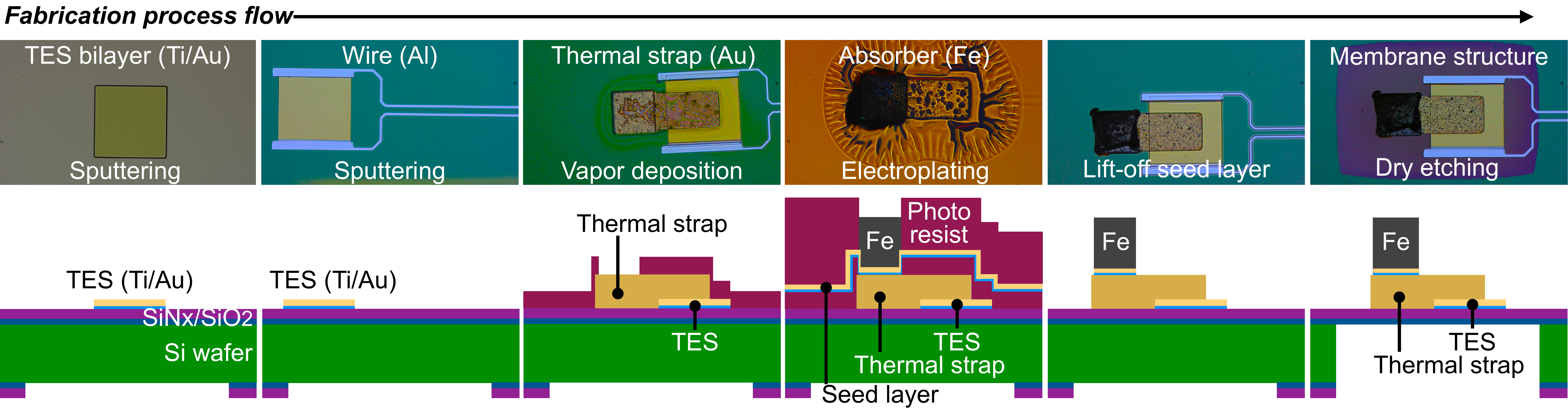}
\caption{Our fabrication process flow of the TES microcalorimeter placed with a distance between the iron absorber and the TES. (Color figure online.)}
\end{center}
\label{fig:3}
\end{figure}

\vspace{-14mm}
\section{Non-uniformity of Iron Absorber and Investigation by Material Analysis}
We observed the fabricated chips and found unexpected black materials 
from the back in all elements after membrane structure formation, as shown in Fig. 4, a.
As the iron surface was rough and curled, we suspected that 
the
irons,
designed only on gold thermal straps, 
were
deposited wrongly.
For analyzing these materials,  
an element
was vertically cut with 
a
focused ion beam (FIB) (Fig. 4, b), 
and we observed the cross-section by 
the
scanning electron microscope (SEM) and energy dispersive X-ray spectroscopy (EDS) (Fig. 4, c, d, e). These images indicate that 
the
iron and the seed layer
curled 
 up together due to the iron tensile stress in an early phase of the 
first electroplating,
and then the gold strap was also peeled off partially from the wafer. In the second electroplating, the iron structures were deposited under the seed layer and 
on and under
the gold strap. Although the size of the gold 
strap
has
a 
5-$\si{\micro m}$
 extra margin
to the iron size not to protrude from the strap,
 if the photolithographic pattern on the 
strap
is out of alignment by more than the margin, a gap between a photoresist and a strap 
edge occurs. 
Therefore,
the
 iron solution could encroach on the back of the gold straps from the gap. This hypothesis indicates that  more accurate pattering or designing the larger strap area than the iron absorber
is
favored. Also, to increase the adhesion between the seed layer and the strap, a new gold 
electroplating
process for straps is under 
testing
to fabricate smoother 
surfaces.
In general, the 
electroplating
 makes 10 times higher thermal conductivity at $\SI{4}{K}$ compared to the vapor deposition
 we used this time.
The
residual resistance ratio 
  (RRR)
of $\SI{300}{K}$ to $\SI{4}{K}$
 is around 30 [\cite{Nagayoshi+20}].

\vspace{-3mm}
\begin{figure}[htbp]
\begin{minipage}{0.59   \hsize} 
\begin{center}
\includegraphics[width=0.85\linewidth, keepaspectratio]{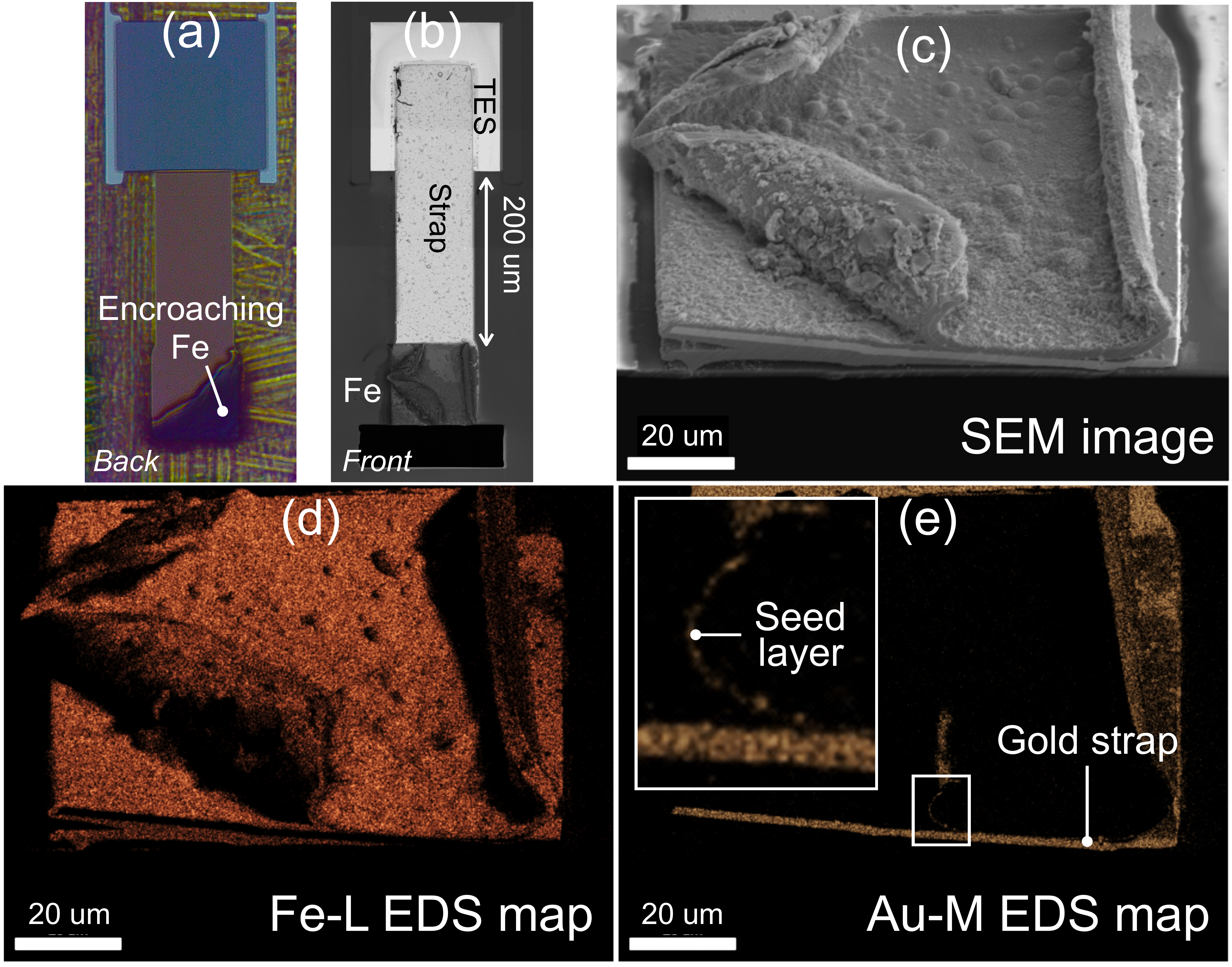}
\end{center}
\caption{(a) Optical microscopical image of calorimeter back side. (b) FIB processing position seen from the front. This element has a 200-$\si{\micro m}$ distance from the iron to the TES. (c) The SEM image of the iron surface and the cross-section. (d) Fe-L EDS map. The iron membrane was deposited under the gold strap, not only on it. (e) Au-M EDS map. The iron membrane and the seed layer curled up together. (Color figure online.)}
\label{fig:4}
\end{minipage}
\hspace{1mm}
\begin{minipage}{0.4\hsize} 
\vspace{-8.1mm}
\begin{center}
\includegraphics[width=1.0\linewidth, keepaspectratio]{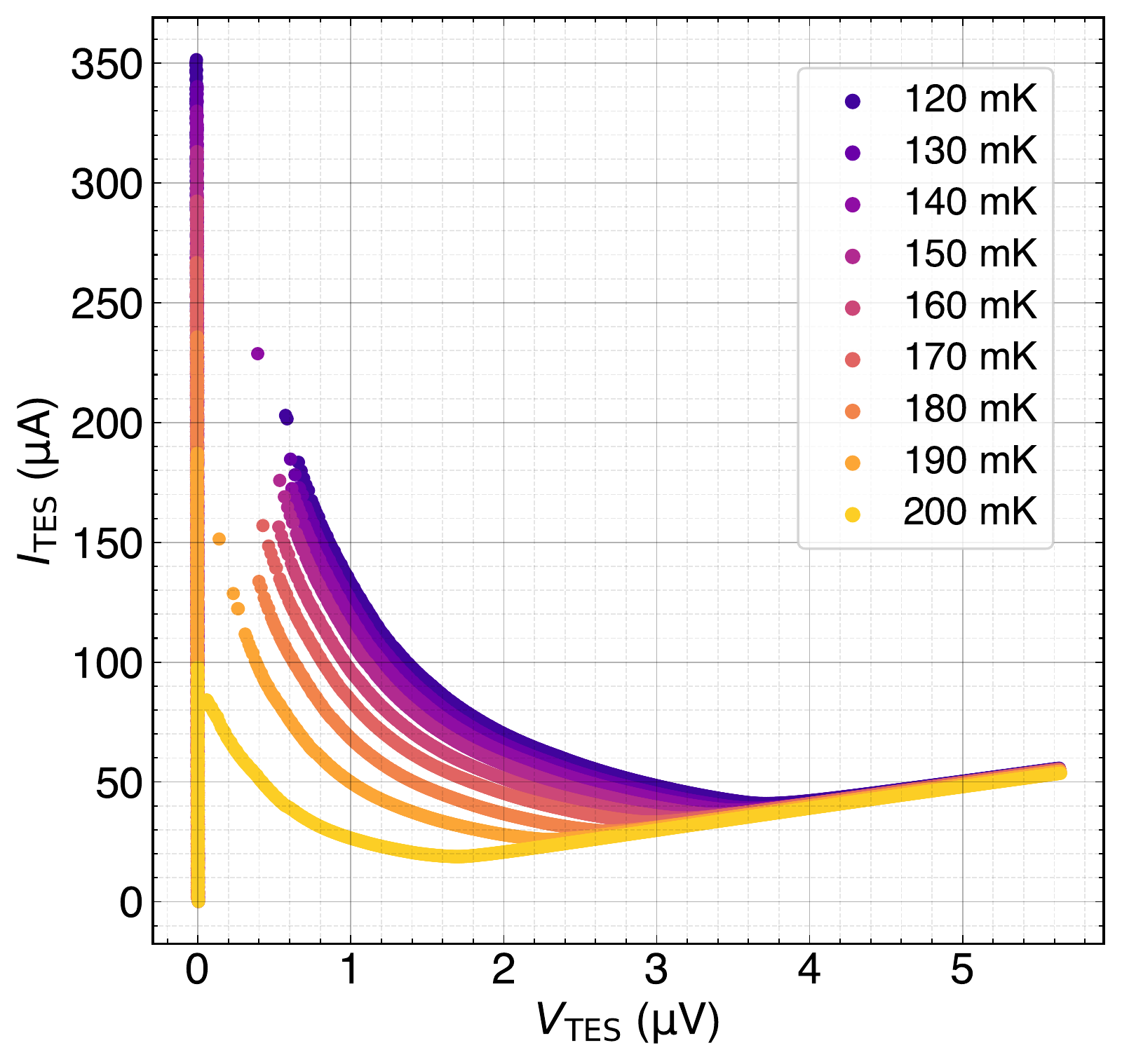}
\end{center}
\caption{Corrected IV, current and voltage, the curve of the TES with a $\num{20}$-$\si{\micro m}$ distance from the iron absorber at different bath temperatures. (Color figure online.)}
\label{fig:5}
\end{minipage}
\end{figure}

\vspace{-7mm}
\section{X-Ray Irradiation Test and Pulse Shapes}
We performed an X-ray irradiation test under 
a
cryogenic temperature of an iron-absorber-TES microcalorimeter with a $\num{20}$-$\si{\micro m}$ distance between the absorber and the TES 
(Fig. \ref{fig:2}).
Figure \ref{fig:5}
shows
the
measured IV curve at different bath temperatures.
An X-ray source ($^{55}$Fe isotope) and a silicon collimator were mounted above the chip at about $\SI{6}{mm}$ and $\SI{320}{\micro m}$, respectively. 
The normal state resistance is $\SI{136}{m\Omega}$.
The transition temperature was $\SI{214}{mK}$, and the
operating
temperature was set to $\SI{130}{mK}$.
The $\alpha$ ($=d\ln{R}/d\ln{T})$ is around 100.
The TES was connected to a superconducting quantum interference device (SQUID) amplifier and operated under 
a standard electro-thermal feedback (ETF) method.

\begin{figure}[htbp]
\begin{minipage}{0.49   \hsize} 
\begin{center}
\includegraphics[width=0.87\linewidth, keepaspectratio]{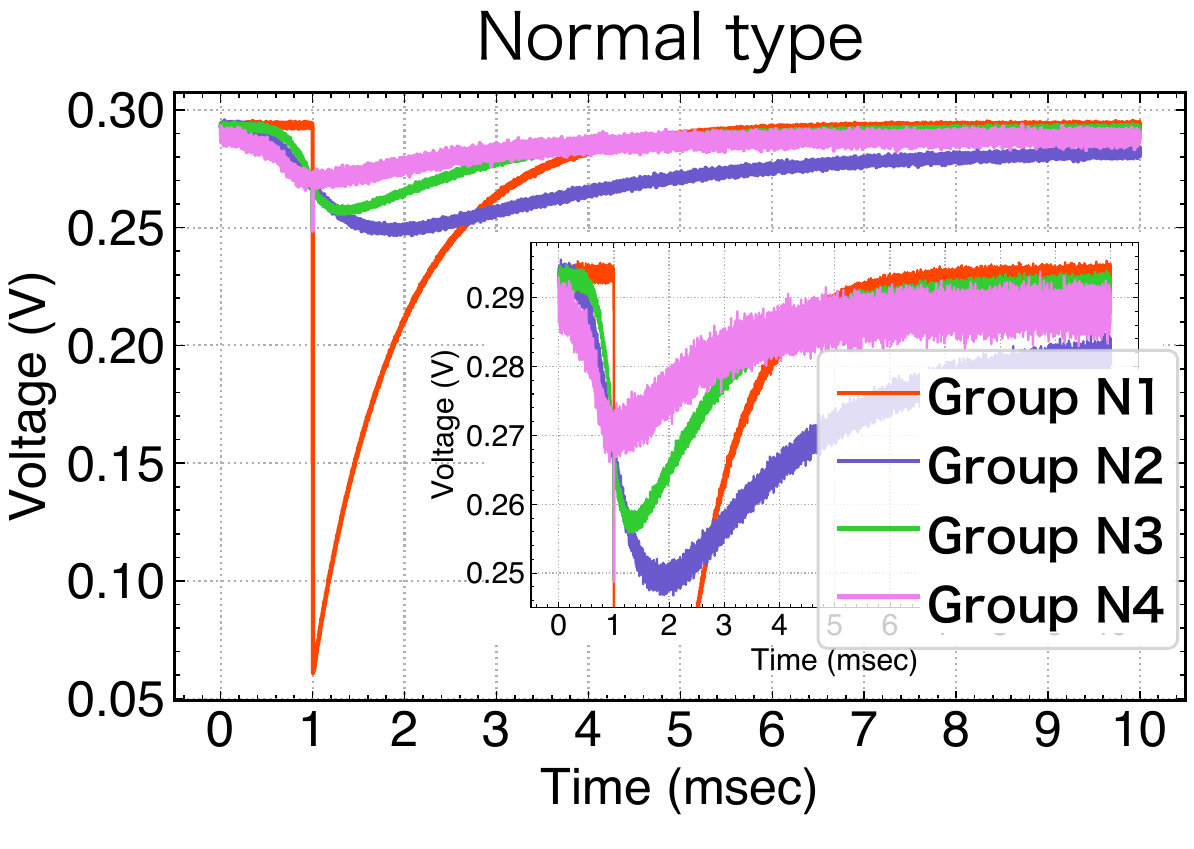}
\end{center}
\end{minipage}
\begin{minipage}{0.49\hsize} 
\begin{center}
\includegraphics[width=0.87\linewidth, keepaspectratio]{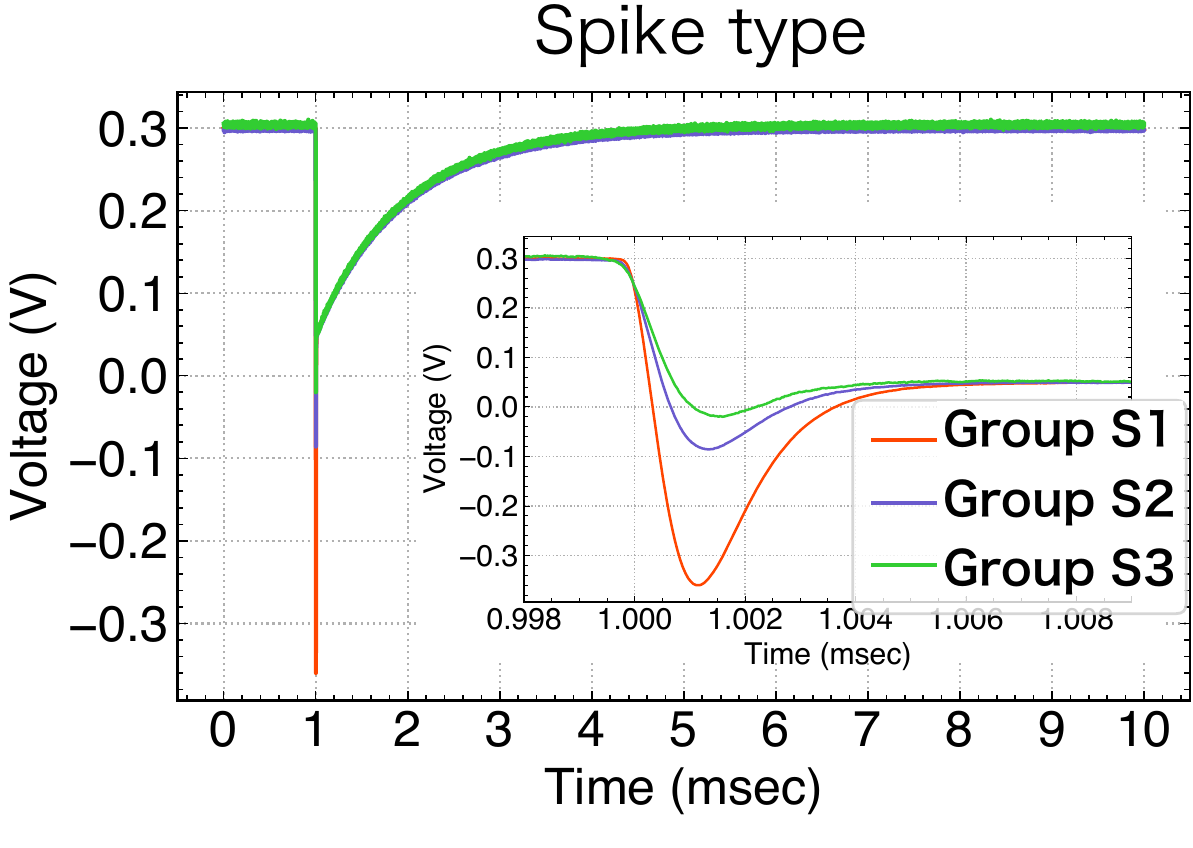}
\end{center}
\end{minipage}
\caption{The average pulses in each group classified by shape. (Color figure online.)}
\label{fig:6}
\end{figure}

In about three days,
698 waveforms of signal pulses were recorded with a time resolution of $\SI{24}{ns}$.  
The pulse shapes 
were
 not identical or similar. 
These were divided
 into two types. 
"Normal" type pulses are characterized by their rising time in the order between microseconds and  milliseconds and by the falling time in a few milliseconds. Others are called "Spike" type pulses because a few microseconds fast spike is seen at the beginning of the pulses,
followed by a slow falling time
in about a millisecond.
The pulse events 
were
 classified into 
 nine
 groups by their pulse shapes, and averaged pulses in each group are shown in Fig. \ref{fig:6}.
The number of pulses, rising time $\tau_{\rm rise}$, and falling time $\tau_{\rm fall}$ 
in each group
 are in Tab. 1.
X-ray radiation spread over
the
vicinity of
the
iron absorber, including
the
thermal strap and TES area, because the 
collimator was not small enough
compared to
the distance from the chip. 
Consequently, 
X-rays
could deposit at the absorber, thermal strap, TES, and membrane and make 
signal pulses with different time
scales.
We are working
on reproducing
obtained pulse shapes with a realistic thermal model
using
a finite element method with COMSOL Multiphysics software [\cite{Mori+21}] to identify the event location by pulse shapes.  
According to the simulation, the iron absorber and the thermal strap
events
have similar pulse shapes. Regardless of absorbing
positions in
the absorber or strap, events up to about $\SI{50}{\micro m}$ far from
the
TES edge have
a
spike shape, and events farther than that have
a
normal shape.
The falling times of both shapes are
$\SI{0.47}{ms}$.
In either of the pulse shapes, the rise time of absorber events is about $\num{0.5}$--$\SI{2.7}{\micro s}$ and is faster and closer to the TES. On the other hand, the thermal strap events under Fe are about the same, and the events closer to TES have an even faster rising time.
They are comparable with Group N1 and S in Tab. \ref{tab:1}. Therefore,
the
obtained data include
both the iron absorber and gold strap events
 at a fraction dependent on the absorption rate of each material.
 As intended,
the X-ray photons deposited on the iron absorber are detected through the thermal strap.

\vspace{-4mm}

\vspace{-9mm}
\begin{table}[htbp]
\caption{The number of pulses, rising time $\tau_{\rm rise}$, and falling time $\tau_{\rm fall}$ in each group. In the spike type, $\tau_{\rm fall}$ is the second slow falling time.}
\centering
\scalebox{0.75}[0.75]{
\begin{tabular}{c|ccccc|cccc}
\hline\hline
\hspace{-1.5mm}Pulse type                & \multicolumn{5}{c|}{Normal type}                                           & \multicolumn{4}{c}{Spike type}           \\ \hline
\hspace{-1mm}Group             & N1        &  N2                & N3                      & N4                      & The others & S1         & S2        & S3        & The others \\
\hspace{-1mm}Number        & 151       &  57               & 117                    & 20                     & 33   & 206       & 80       & 18       &16     \\
$\tau_{\rm rise}$ (${\rm \mu s}$) & 0.40--1.7 & 4.0$\times 10^2$ & (1.0--3.0)$\times 10^2$ &  (1.3--2.5)$\times 10^3$ & -     & 0.70--0.90 & 1.0--1.2  & 1.2--1.4  & -      \\
$\tau_{\rm fall}$ (ms)          & 0.80--0.90 & 4.0--5.3          &  1.2--2.2                & 0.10--0.30              & -      & 0.80--1.0  & 0.90--1.0 & 0.90--1.0 & -      \\ \hline\hline
\end{tabular}
}\label{tab:1}
\end{table}

\vspace{-7 mm}
\section{Conclusion and Future Prospects}
We successfully deposited iron absorbers whose thermal conductivities are 
$\num{2.0}$--$\SI{5.4}{W/K/m}$ at $\SI{4}{K}$ using $^{56}$Fe and produced 
the
first test calorimeters
for the solar axion detector.
In irradiation tests, it was likely that iron events 
were
detected. 
The identification of all events still needs to be completed. However,  for the first time,
we successfully operated the TES microcalorimeter with a certain distance 
between
the
iron
absorber
and
the TES under iron magnetization. 
 The
high thermal conductivity of gold thermal straps could decrease the effect of degrading the energy resolution by the position dependence in an absorber discussed in [\cite{Konno+20}]. To make efficient detectors, we try to deposit gold straps by electroplating and
continue to
fabricate iron absorbers with high 
conductivity.

\begin{acknowledgements}
We are grateful to S. Moriyama for his suggestions about solar axion searches. 
Our research was partially performed in the nano-electronics clean room
at the
Institute of Space and Astronautical Institute (ISAS).
 We also thank 
 the
 technical staff
  members of
 the Nanotechnology Research Center (NTRC)
 at
 Waseda University and 
 the
 National Institute for Materials Science (NIMS). The SQUID array amplifier
 used
 for irradiation tests was fabricated by the clean room for analog \& digital superconductivity (CRAVITY) of 
 the
 National Institute of Advanced Industrial Science and Technology (AIST). 
 The
 Japan Society financially supported this work for the Promotion of Science (JSPS) KAKENHI Grant Number 18H01244 and 20H05857.
\end{acknowledgements}

\pagebreak

\end{document}